\documentclass[preprint2, trackchanges]{aastex63}
\usepackage{multirow}
\usepackage{amsmath}
\usepackage{appendix}
\usepackage{CJK}

\begin{document}
\begin{CJK*}{UTF8}{gbsn}

\title{Understanding the Lateral Drifting of an Erupting Filament with a Data-constrained Magnetohydrodynamic Simulation}

\correspondingauthor{P. F. Chen and Y. Guo}
\email{chenpf@nju.edu.cn, guoyang@nju.edu.cn}

\author[0000-0002-4205-5566]{J. H. Guo (郭金涵)}
\affiliation{School of Astronomy and Space Science and Key Laboratory for Modern Astronomy and Astrophysics, Nanjing University, Nanjing 210023, China}

\affiliation{Centre for Mathematical Plasma Astrophysics, Department of Mathematics, KU Leuven, Celestijnenlaan 200B, B-3001 Leuven, Belgium}

\author[0000-0002-1190-0173]{Y. Qiu (邱晔)}
\affiliation{School of Astronomy and Space Science and Key Laboratory for Modern Astronomy and Astrophysics, Nanjing University, Nanjing 210023, China}

\author[0000-0002-9908-291X]{Y. W. Ni (倪仪伟)}
\affiliation{School of Astronomy and Space Science and Key Laboratory for Modern Astronomy and Astrophysics, Nanjing University, Nanjing 210023, China}

\author[0000-0002-9293-8439]{Y. Guo (郭洋)}
\affiliation{School of Astronomy and Space Science and Key Laboratory for Modern Astronomy and Astrophysics, Nanjing University, Nanjing 210023, China}

\author[0000-0001-7693-4908]{C. Li (李川)}
\affiliation{School of Astronomy and Space Science and Key Laboratory for Modern Astronomy and Astrophysics, Nanjing University, Nanjing 210023, China}

\author[0000-0002-6641-8034]{Y. H. Gao (高宇航)}
\affiliation{School of Earth and Space Sciences, Peking University, Beijing 100871, China}
\affiliation{Centre for Mathematical Plasma Astrophysics, Department of Mathematics, KU Leuven, Celestijnenlaan 200B, B-3001 Leuven, Belgium}

\author[0000-0003-3364-9183]{B. Schmieder}
\affiliation{Centre for Mathematical Plasma Astrophysics, Department of Mathematics, KU Leuven, Celestijnenlaan 200B, B-3001 Leuven, Belgium}
\affiliation{LESIA, Observatoire de Paris, CNRS, UPMC, Universit\'{e} Paris Diderot, 5 place Jules Janssen, F-92190 Meudon, France}

\author[0000-0002-1743-0651]{S. Poedts}
\affiliation{Centre for Mathematical Plasma Astrophysics, Department of Mathematics, KU Leuven, Celestijnenlaan 200B, B-3001 Leuven, Belgium}
\affiliation{Institute of Physics, University of Maria Curie-Sk{\l}odowska, ul.\ Radziszewskiego 10, 20-031 Lublin, Poland}

\author[0000-0002-7289-642X]{P. F. Chen (陈鹏飞)}
\affiliation{School of Astronomy and Space Science and Key Laboratory for Modern Astronomy and Astrophysics, Nanjing University, Nanjing 210023, China}

\begin{abstract}
	Solar filaments often exhibit rotation and deflection during eruptions, which would significantly affect the geoeffectiveness of the corresponding coronal mass ejections (CMEs). Therefore, understanding the mechanisms that lead to such rotation and lateral displacement of filaments is a great concern to space weather forecasting. In this paper, we examine an intriguing filament eruption event observed by the Chinese H$\alpha$ Solar Explorer (CHASE) and the Solar Dynamics Observatory (SDO). The filament, which eventually evolves into a CME, exhibits significant lateral drifting during its rising. Moreover, the orientation of the CME flux rope axis deviates from that of the pre-eruptive filament observed in the source region. To investigate the physical processes behind these observations, we perform a data-constrained magnetohydrodynamic (MHD) simulation. Many prominent observational features in the eruption are reproduced by our numerical model, including the morphology of the eruptive filament, eruption path, and flare ribbons. The simulation results reveal that the magnetic reconnection between the flux-rope leg and neighboring low-lying sheared arcades may be the primary mechanism responsible for the lateral drifting of the filament material. Such a reconnection geometry leads to flux-rope footpoint migration and a reconfiguration of its morphology. As a consequence, the filament material hosted in the flux rope drifts laterally, and the CME flux rope deviates from the pre-eruptive filament. This finding underscores the importance of external magnetic reconnection in influencing the orientation of a flux rope axis during eruption.

\end{abstract}

\keywords{Magnetohydrodynamical simulations (1966); Solar coronal mass ejections (310); Solar filaments (1495); Solar magnetic fields (1503); Solar flares (1496)}

\section{Introduction} \label{sec:intro}

Coronal mass ejections (CMEs) are large-scale transient eruptions that occur in the solar corona, expelling substantial amounts of magnetized plasma into the interplanetary space and driving disturbances to the heliosphere environment \citep{Chen2011}. As they propagate through the interplanetary space, they are termed as interplanetary CMEs (ICMEs). It is believed that the core magnetic structure of a CME/ICME is a magnetic flux rope (albeit not always), characterized by helical field lines winding around one axis. Flux ropes can be detected through a variety of remote sensing observations in the low corona, such as filaments, sigmoids, coronal cavities, and hot channels \citep{Cheng2017}. In-situ measurements of the interplanetary plasma can identify flux ropes through large and smooth rotations of magnetic fields, which are typically referred to as magnetic clouds \citep[MCs;][]{Burlaga1981}. It is well accepted that the orientation of a flux-rope axis is crucial for predicting adverse space weather events, and the southward magnetic-field component carried by the flux rope is the most dominant indicator for the possible generation of geomagnetic storms \citep{Liu2016, Kilpua2019, Shen2022, Tsurutani2022}, which are triggered by the reconnection between the northward geomagnetic field and the southward interplanetary magnetic field.

Solar filaments are often observed as the progenitors of CMEs, for instance, over 70\% of all CMEs are associated with eruptive filaments \citep{Munro1979}. On the other hand, filaments are prominent proxies for flux ropes in the low corona, as statistical research has shown that 89\% of the eruptive filaments are supported by flux ropes prior to eruption \citep{Ouyang2017}. Studies have also shown a strong correlation between the magnetic structures of MCs and their pre-eruptive filaments in the source region. \citet{Bothmer1994} found that in four out of their five cases, the orientation and chirality of MCs are the same as those of the pre-eruptive filaments. \citet{Yurchyshyn2001} demonstrated that the axial direction and helicity of the MC magnetic fields are in alignment with those of the corresponding eruptive filaments. Furthermore, \citet{Wang2006} found that the tilt angles of some MCs of their sample are nearly parallel with the corresponding filaments, particularly for those whose flux-rope axes are almost perpendicular to the Sun-Earth line. As such, the determination of the magnetic morphology of solar filaments provides valuable insights into the magnetic structures of the resulting ICMEs/MCs, making it possible to predict the orientation of interplanetary magnetic fields near the Earth or even the possibility of geomagnetic storms by examining solar filaments.

However, we cannot always expect that a filament erupts in the radial direction, keeping its orientation all the way from the low corona to the interplanetary space. As a result, inconsistency between the original filament and the resulting MC axis was reported \citep{Wang2006}. The mechanisms that lead to such inconsistencies fall into two categories, i.e.\ the deflection and the rotation of the flux rope axis. For the former mechanism, it would alter the original trajectory and guide some non-Earth-directed CMEs eventually to arrive at the Earth or vice versa \citep{Wang2002, Wang2004, Wang2011}. Moreover, \citet{Gopalswamy2014} found that some CMEs erupting from the backside of the Sun or behind the limb can also result in strong solar energetic particles and ground level enhancement events, suggesting that there should exist deflection causing the CMEs to deviate from the original trajectory. Different from deflection, the rotation of a flux rope usually occurs in the early stage of a CME. For example, many filaments frequently undergo rotations as they rise continuously \citep{Green2007, Zhou2019, Zhou2020}, leading to variations of their axis orientation. \citet{Song2018} reported a very intriguing prominence eruption, which undergoes counterclockwise and clockwise rotations sequentially. \citet{Shiota2010} found that magnetic reconnection can lead to the rotation of the CME flux rope. Additionally, \citet{Liuy2018} found that the axis of a CME flux rope deviates from the initial orientation by about $95^{\circ}$. All of these studies indicated that the deflection and rotation of a flux rope commonly occur during propagation in the corona and interplanetary space, which may greatly influence its geoeffectiveness. Among them, the deflection influences whether a CME arrives at the Earth, and the rotation could alter the $B_{z}$ profile of an ICME, in particular its duration and strength of the southern component, which significantly affects its magnetic reconnection with the northward geomagnetic field \citep{Maharana2023}. Therefore, uncovering the underlying physical processes behind these observational phenomena is crucial for refining CME forecasting models.

To unveil the physical mechanisms of the flux rope deflection and axis rotation, data-driven or data-constrained numerical simulations are fairly powerful tools. These simulations utilize the observational data, such as the magnetic field, velocity field, and electric fields on the photosphere, to derive the initial condition for the numerical setup and constrain the subsequent evolution \citep{Jiang2022}. Such models have played a tremendous role in reproducing the three-dimensional (3D) evolution of the magnetic fields \citep{Cheung2012, Jiang2016, Inoue2018a, Inoue2018b, Guo2019, Pomoell2019, Guoy2021} and the thermodynamics \citep{Torok2018, Fan2022, Guo2023} in the corona, as well as the ICME propagation in the interplanetary space \citep{Pomoell2018, Stefaan2020, Scolini2020, Verbeke2022}. Here, we perform a nonadiabatic data-constrained MHD simulation of the filament eruption event that occurred on 2022 August 18, which exhibits obvious lateral drifting during its ascent, as captured by SDO \citep{Pesnell2012} and CHASE \citep{Lic2019, Lic2022} simultaneously. One aim is to reproduce as many observational characteristics as possible in the simulation. Additionally, we expect to explain why this filament displays lateral drifting during the eruption and explore how it affects the magnetic structure of the resulting CME. We describe the observations in Section~\ref{sec:obs}, introduce the setup of the MHD modeling in Section~\ref{sec:met}, and present the simulation results in Section~\ref{sec:res}, which are followed by the summary and discussion in Section~\ref{sec:dis}.

\section{Multi-wavelength observations}\label{sec:obs}
\begin{figure*}[htbp]
	\includegraphics[width=15cm,clip]{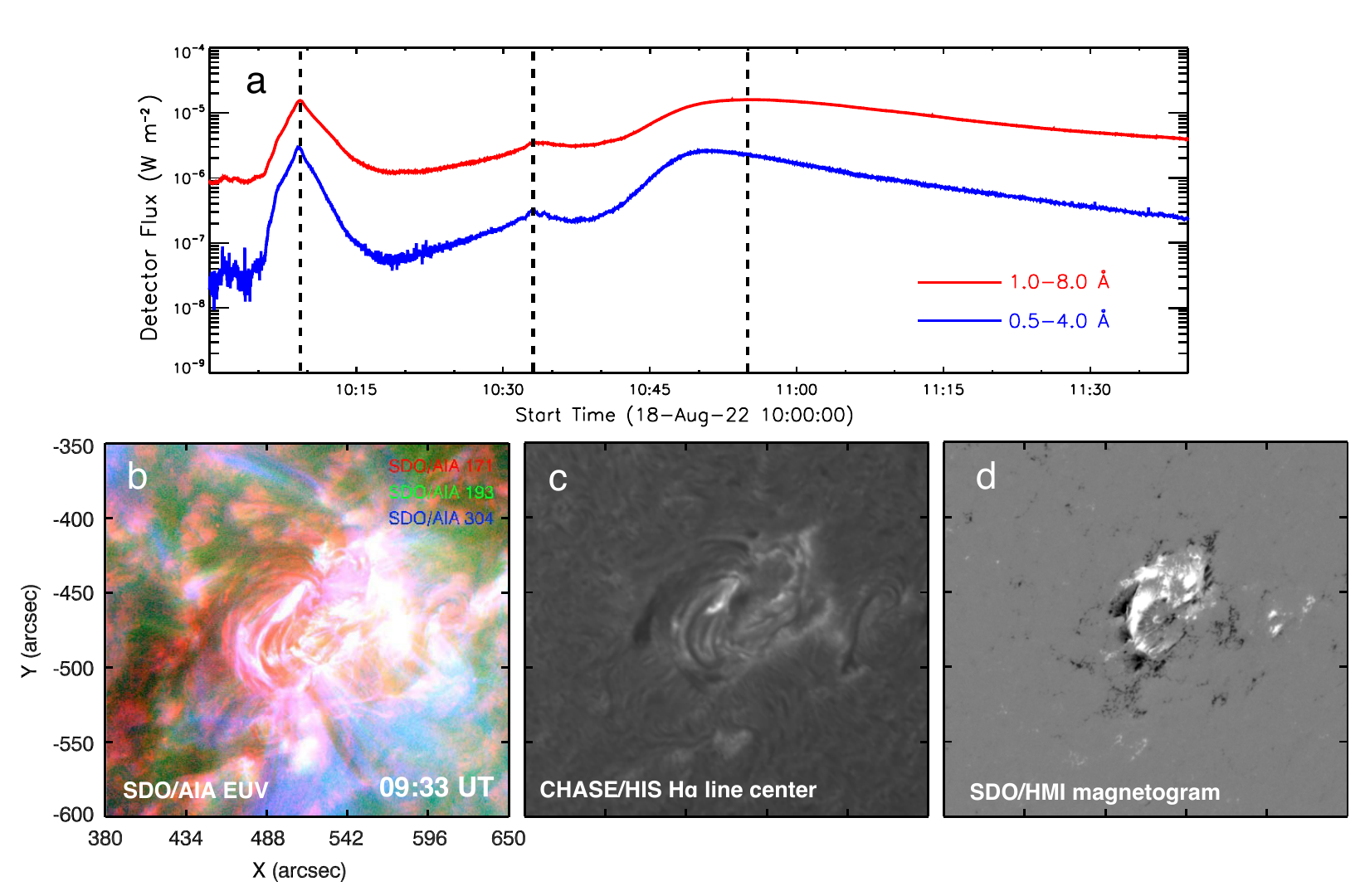}
	\centering
	\caption{(a) GOES SXR light curves between 10:00~UT and 11:40~UT on 2022 August 18, in which the peak times are marked with dashed lines. (b) Composite image of AIA 171~\AA\ (red), 193~\AA\ (green) and 304~\AA\ (blue) of the pre-eruptive filament at 09:33~UT on 18 August 2022. (c) H$\alpha$ line-center image of the pre-eruptive filament observed by CHASE/HIS at 09:33~UT. (d) Line-of-sight magnetic fields (gray scale) in the source region of the pre-eruptive filament. \label{fig1}}
\end{figure*}

\begin{figure*}[htbp]
	\includegraphics[width=13cm,clip]{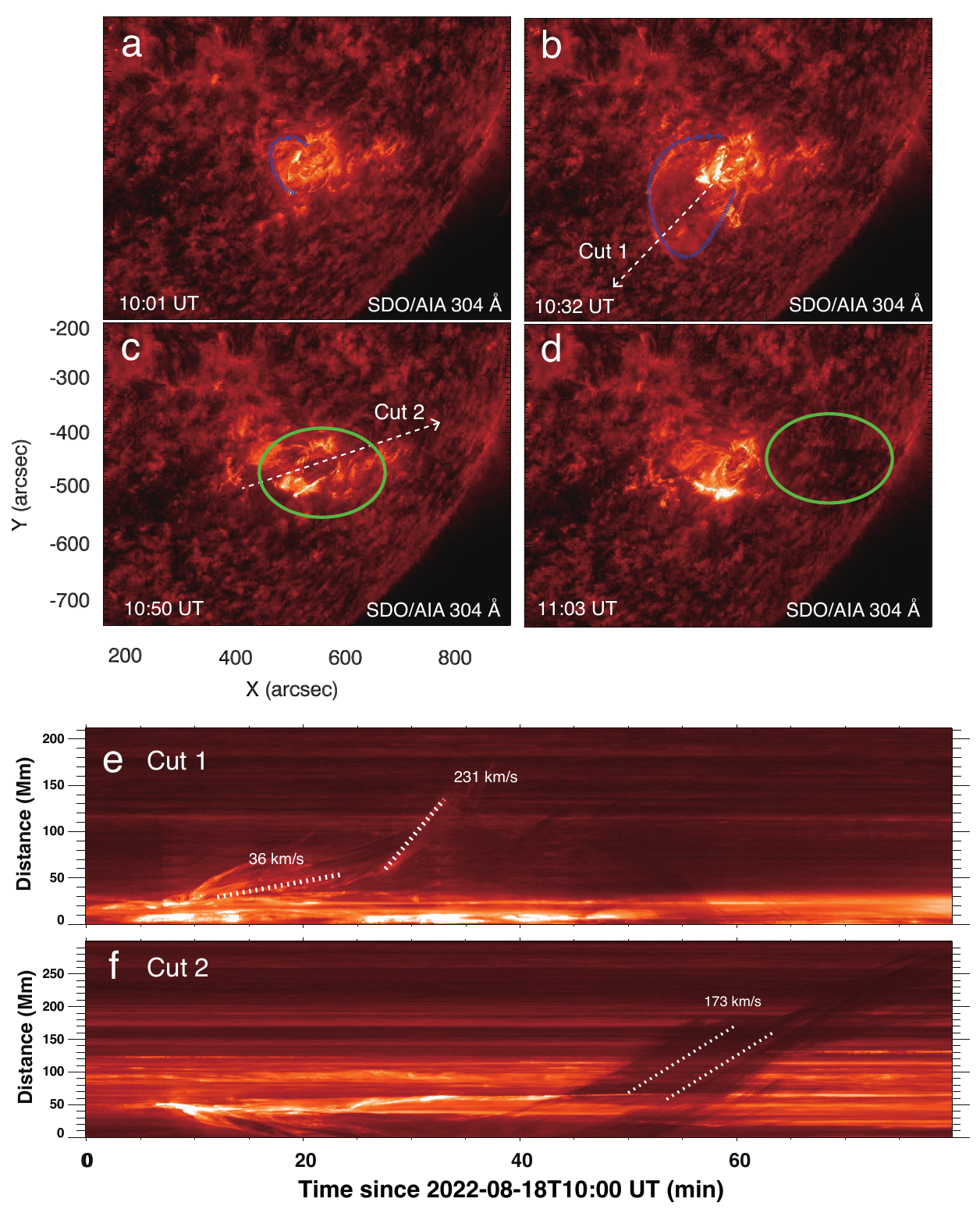}
	\centering
	\caption{SDO/AIA 304~\AA\ observations showing the evolution of the filament. The blue dashed lines in panels~(a--b) depict the morphology of the filament during eruption. The green circles in panels~(c--d) decipher the non-radial motion of the filament material. panels~(e) and (f) show the stack plots of the 304~\AA\ intensity along Cut 1 and Cut 2 in panels~(b) and (c), respectively. (The animation shows the filament eruption process from 10:00 UT to 11:20 UT. (An animation of this figure is available.)} \label{fig2}
\end{figure*}

The event under study is the filament eruption that occurred at 10:00~UT on 2022 August 18, in NOAA active-region 13078. Figure~\ref{fig1}a displays the GOES soft X-ray (SXR) light curves in 0.5--4 and 1−-8~\AA\ during 10:00--11:40~UT. It is found that this filament eruption is accompanied by an M1.3 flare (10:00-10:13~UT), a C3.4 flare (10:19-10:37~UT), and an M1.5 flare (10:37-11:13~UT). Figures~\ref{fig1}b--\ref{fig1}d displays the multi-wavelength observations of the pre-eruptive filament observed by SDO/Atmospheric Imaging Assembly \citep[AIA;][]{Lemen2012} and CHASE/H$\alpha$ Imaging Spectrograph (HIS) \footnote{https://ssdc.nju.edu.cn}, and the magnetic fields in its source region at 09:33~UT on 18 August 2022 observed by SDO/Helioseismic and Magnetic Imager \citep[HMI;][]{Scherrer2012}. It is seen that the filament has a crescent shape and lies along the polarity inversion line on the east side of the active region. The orientation of the filament spine is almost from north to south. The eruption of this filament can be divided into two distinct stages: the ejection stage and the subsequent lateral drifting stage (Animation of Figure~\ref{fig2}). At 10:00~UT, the filament starts to rise slowly and leans toward the southeast with a semi-circle morphology, as shown in Figures~\ref{fig2}a and \ref{fig2}b. Hereafter, the entire bulk of the eruptive filament is no longer discernible as before, and a cloud of faint filament material drifts to the west (outlined by green ovals in Figures~\ref{fig2}c--\ref{fig2}d). To show the filament dynamics clearly, we select two cuts (Cut 1 in Figure~\ref{fig2}b and Cut 2 in Figure~\ref{fig2}c) along the eruption direction and drifting direction to make time-distance maps (Figure~\ref{fig2}e and \ref{fig2}f). One can see that the rising of this erupting filament falls into two stages (Figure~\ref{fig2}e): in the beginning, the filament rises with a relatively low velocity (36~km s$^{-1}$), and then accelerates and rises with a higher mean velocity of about 231~km s$^{-1}$. As for the filament drifting, it is seen that the filament material starts to drift westward at around 10:40~UT, with a mean velocity of about 173~km s$^{-1}$ (Figure~\ref{fig2}f). Intriguingly, these three stages can be distinguished very well by three flares, implying that magnetic reconnection may play an important role in the eruption and drifting of this filament. Moreover, it is observed that the drifting direction of the filament is nearly opposite to its initial eruption direction. What mechanisms lead to such a lateral drifting of the filament material during the eruption? We try to answer this question in the following sections of this paper.

\begin{figure*}[htbp]
	\includegraphics[width=16cm,clip]{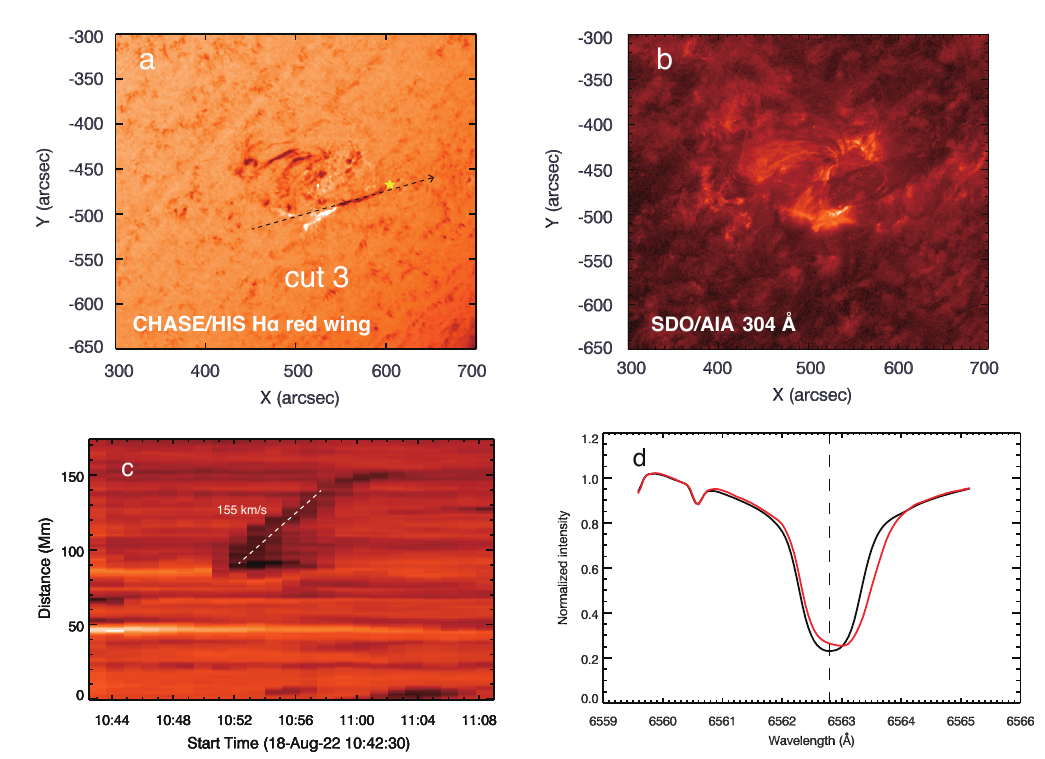}
	\centering
	\caption{(a) CHASE/HIS observation in H$\alpha$ red-wing image at 10:55~UT, where the black dashed line (Cut 3)~represents the cut to make the time-distance map in panel~(c), and the yellow asterisk represents the point to show the H$\alpha$ spectrum profile in panel~(d). (b) SDO/AIA 304~\AA\ image at 10:55~UT. (c) Time-distance diagram along Cut 3 in panel~(a). (d) H$\alpha$ spectrum profile (red solid line) at the asterisk in panel~(a). The black solid line represents the reference profile observed from the quiet Sun, and the vertical dashed line is the reference line center (6562.8 \AA). \label{fig3}}
\end{figure*}

The CHASE/HIS H$\alpha$ spectral observations come with much more information about the lateral drifting of the filament material. On the one hand, the H$\alpha$ spectral line is highly sensitive to cold plasmas, such as the filament and chromospheric activities, making it more suitable than extreme-ultraviolet (EUV) wavebands in identifying the fine structures of filaments. On the other hand, CHASE/HIS provides full-disk H$\alpha$ spectroscopic observations. As such, the slit-jaw images at different wavelengths and Doppler velocity fields can be derived, enabling more precise characterizations of the dynamics of the filament eruption. Unfortunately, the CHASE observatory has not been able to provide continuous observations so far, and only captures the subsequent lateral drifting stage of this filament eruption. In spite of this, the CHASE observations still provide valuable information for this study, enabling the exploration of the underlying mechanisms that lead to the filament lateral drifting. Figure~\ref{fig3}a shows the H$\alpha$ red-wing image (6562.82+0.87 \AA) at 10:55~UT. It is evident that a long, dark structure moving westward in the H$\alpha$ red-wing image is closely aligned with the drifting filament material observed in the SDO/AIA 304~\AA\ image (see Figures~\ref{fig3}b). To examine its transverse movements, in Figure~\ref{fig3}c, we make a time-distance diagram along Cut~3 marked in Figure~\ref{fig3}a. The transverse velocity of this structure is measured to be approximately 155~km s$^{-1}$, which is comparable to the velocity in the SDO/AIA 304~\AA\ images (Figure~\ref{fig2}f). This indicates that the westward movement of the dark structure observed in the H$\alpha$ red-wing waveband corresponds to the westward drift of a bulk of filament material in the SDO/AIA 304~\AA\ images. The spectral profile in Figure~\ref{fig3}d suggests that the spectrum is red-shifted, indicating the downward plasma flows.

\begin{figure*}[htbp]
    \includegraphics[width=15cm,clip]{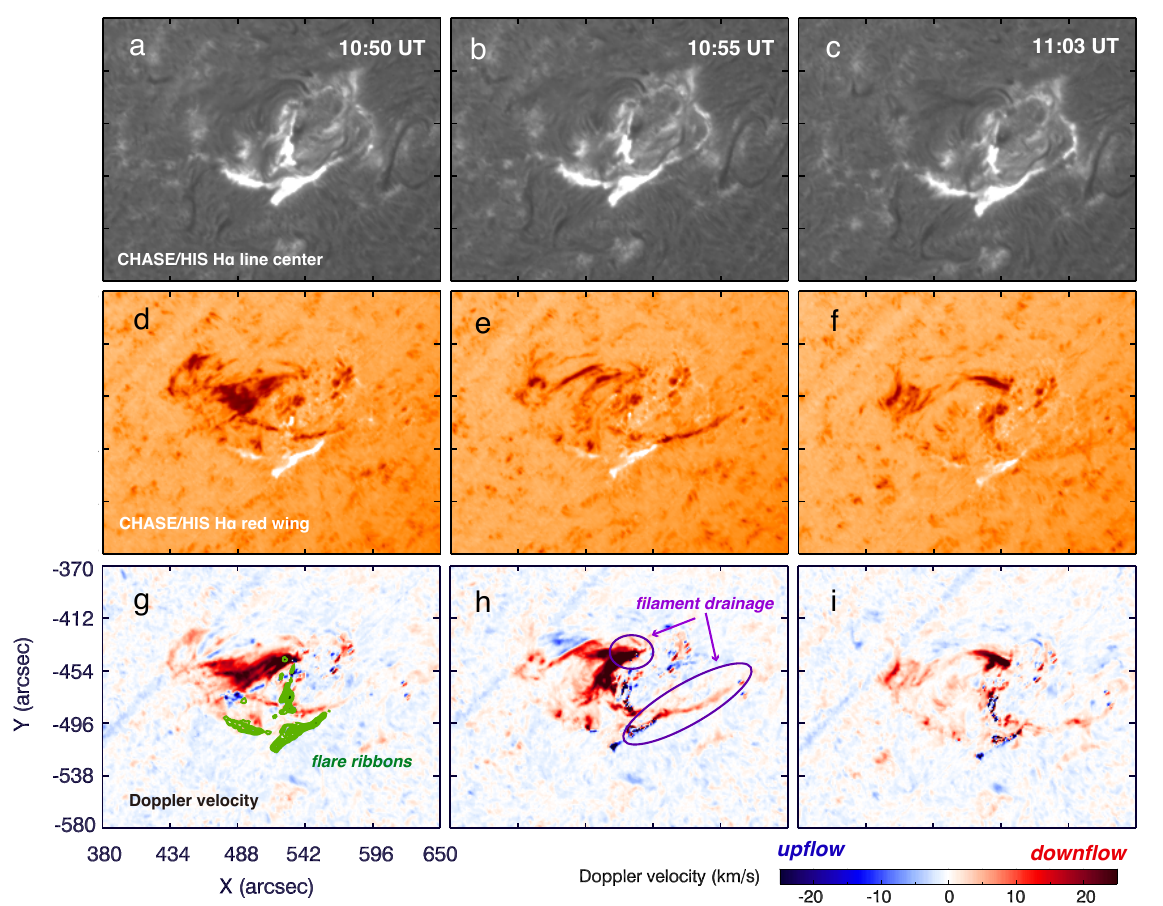}
    \centering
	\caption{CHASE/HIS observations in H$\alpha$ line center (panels a--c), H$\alpha$ red-wing (panels d--f), and the derived velocity maps (panels g--i). The green contours in panel~(g) represent the flare ribbons observed in H$\alpha$ line center image at the same moment. \label{fig4}}
\end{figure*}

Figure~\ref{fig4} displays the images of CHASE/HIS H$\alpha$ line center, H$\alpha$ red-wing, and Doppler velocity fields from the top to bottom, respectively, where the velocity fields are calculated with the moment method. \citet{Qiu2022} provides more details on the calibration procedures of the CHASE data. The arc-shaped ribbon observed in SDO/AIA 304~\AA\ can also be clearly seen in the H$\alpha$ line center images, but the westward drifting of a bulk of filament material is absent in the H$\alpha$ line center images. However, it is markedly distinguishable in the H$\alpha$ red wing images (Figures~\ref{fig4}d--\ref{fig4}f) and the Doppler velocity maps (Figures~\ref{fig4}g--\ref{fig4}i). The Doppler velocity map suggest that the laterally drifting filament materials are also falling with downward velocities exceeding 20~km s$^{-1}$, forming conjugate drainage sites, as shown in Figure~\ref{fig4}h. The northern filament drainage site is almost co-spatial with the northern endpoint of the filament. However, the southern drainage site is not around the other endpoint of the filament. Instead, it is located at a negative polarity, which is much more diffuse and further west. The CHASE observations provide a precise characterization of the lateral drifting of the filament material with the 3D velocity field: the westward drifting is accompanied by downward draining. Such a combination of the imaging and spectroscopic observations has also been used to infer the full velocity of the CME bulk motion \citep{Xu2022}. It is seen in the following sections that the recognized drainage sites and their spatial relationships with the photospheric magnetic fields reflect the magnetic structures of its supporting flux rope in the eruption.

\begin{figure*}[htbp]
    \includegraphics[width=15cm,clip]{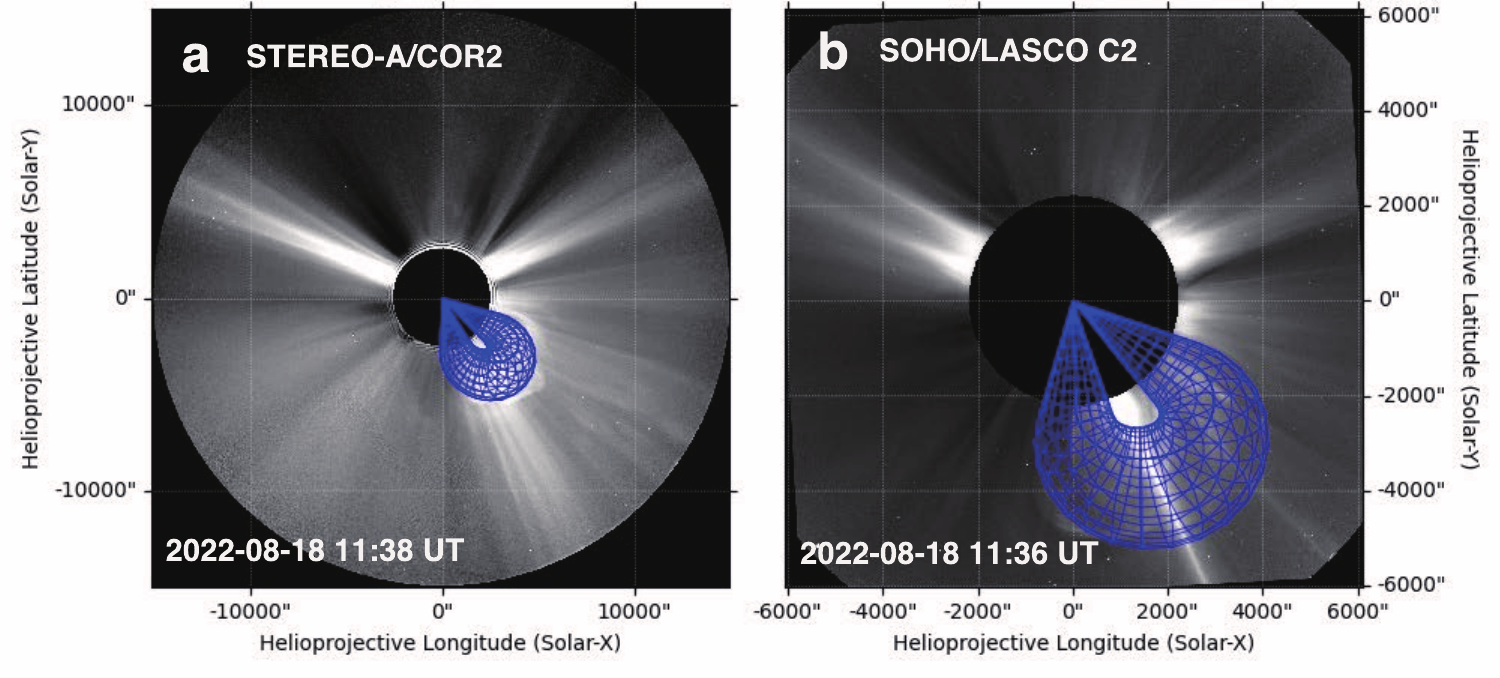}
    \centering
	\caption{GCS reconstruction for the CME observed by the white-light coronagraphs of (a) STEREO A/COR2 and (b) SOHO/LASCO C2. The blue wireframe depicts the CME flux rope structure. \label{fig5}}
\end{figure*}

Subsequently, a CME is observed by two white-light coronagraphs within a time window of two hours, i.e.,  COR2 coronagraph on board the Solar Terrestrial Relations Observatory \citep[STEREO;][]{Kaiser2008} and Large Angle Spectrometric Coronagraph (LASCO) C2 on board the Solar and Heliospheric Observatory \citep[SOHO;][]{Domingo1995}, as shown in Figures~\ref{fig5}a and \ref{fig5}b. To further quantify the structure of the CME flux rope, we implement a graduated cylindrical shell (GCS) reconstruction \citep{Thernisien2011} using the simultaneous white-light observations of this CME from two vantage points. The GCS reconstruction is an empirical model that constructs the 3D structures of CME flux ropes, assuming that the flux rope is in a croissant-like shape with conical legs anchored to the solar surface. It involves six free parameters: the CME apex height ($h$), half-angle width ($\alpha$), aspect ratio ($\kappa$), tilt angle ($\gamma$), longitude ($\phi$) and latitude ($\theta$) of the CME flux rope. Figure~\ref{fig5} displays the fitting results for this CME. It can be seen that the wireframe used to render the constructed flux rope coincides with the CME in the white-light images. The fitted parameters are as follows: $h=6.39\;R_{\odot}$, $\alpha=25^{\circ}$, $\kappa=0.27$, $\gamma=31.5^{\circ}$, $\phi=40^{\circ}$, and $\theta= -46.5^{\circ}$. The reconstruction results suggest that this CME is associated with the eruptive filament depicted in Figure~\ref{fig2}. On the one hand, the latitude and longitude of the CME fall within a range of $\pm 10^{\circ}$ of the pre-eruptive filament in the source region ($\phi=35^{\circ}$ and $\theta=-30^{\circ}$). Additionally, the CME appears in the coronagraph within a time window of two hours starting from the filament eruption. These results are consistent with the common criteria used to check the association between CMEs and filament eruptions \citep{Jing2004}. Moreover, the CME flux rope axis is tilted about $31.5^{\circ}$ with respect to the solar equator. Compared to the pre-eruptive filament observed in the source region, the direction of the flux rope axis has deviated by about $90^{\circ}$ when propagating from the solar surface to 6.39$\;R_{\odot}$.

\section{MHD modeling}\label{sec:met}

To perform a simulation of this eruptive event, we adopt a nonadiabatic MHD model that takes into account the field-aligned thermal conduction, and the governing equations are as follows:

\begin{eqnarray}
 && \frac{\partial \rho}{\partial t} +\nabla \cdot(\rho \boldsymbol{v})=0,\label{eq1}\\
 && \frac{\partial (\rho \boldsymbol{v})}{\partial t}+\nabla \cdot(\rho \boldsymbol{vv}+p_{_{tot}}\boldsymbol{I}-\frac{\ \boldsymbol{BB}}{\mu_{0}})=\rho \boldsymbol{g},\label{eq2}\\
 && \frac{\partial \boldsymbol{B}}{\partial t} + \nabla \cdot(\boldsymbol{vB-Bv})=0,\label{eq3}\\
 && \frac{\partial \varepsilon}{\partial t}+\nabla \cdot(\varepsilon \boldsymbol{v}+p_{_{tot}}\boldsymbol{v}-\frac{\boldsymbol{BB}}{\mu_{0}}\cdot \boldsymbol{v}) =\rho \boldsymbol{g \cdot v} \\ \nonumber
 && +\nabla \cdot(\boldsymbol{\kappa} \cdot \nabla T), \label{eq4}
\end{eqnarray}
where $p_{_{tot}} \equiv p + B^2 / (2\mu_{0})$ is the sum of the thermal pressure and magnetic pressure, $\boldsymbol{g}=-g_{\odot}r_{\odot}^2/(r_{\odot}+z)^2\boldsymbol{e_{z}}$ is the gravitational acceleration, $g_{\odot}= \rm 274\ m\ s^{-2}$ is the gravitational acceleration at the solar surface, $r_{\odot}$ is the solar radius, $\varepsilon =\rho v^2/2+p/(\gamma -1)+ B^2 / (2\mu_{0})$ is the total energy density,  $\nabla \cdot(\boldsymbol{\kappa} \cdot \nabla T)=\nabla \cdot(\kappa_{\parallel}\boldsymbol{\hat{b}\hat{b}} \cdot \nabla T)$ represents field-aligned thermal conduction, and $\kappa_{\parallel} =10^{-6}\ T^{\frac{5}{2}}\ \rm erg\ cm^{-1}\ s^{-1}\ K^{-1}$ is the Spitzer heat conductivity.

\begin{figure*}[htbp]
    \includegraphics[width=13cm,clip]{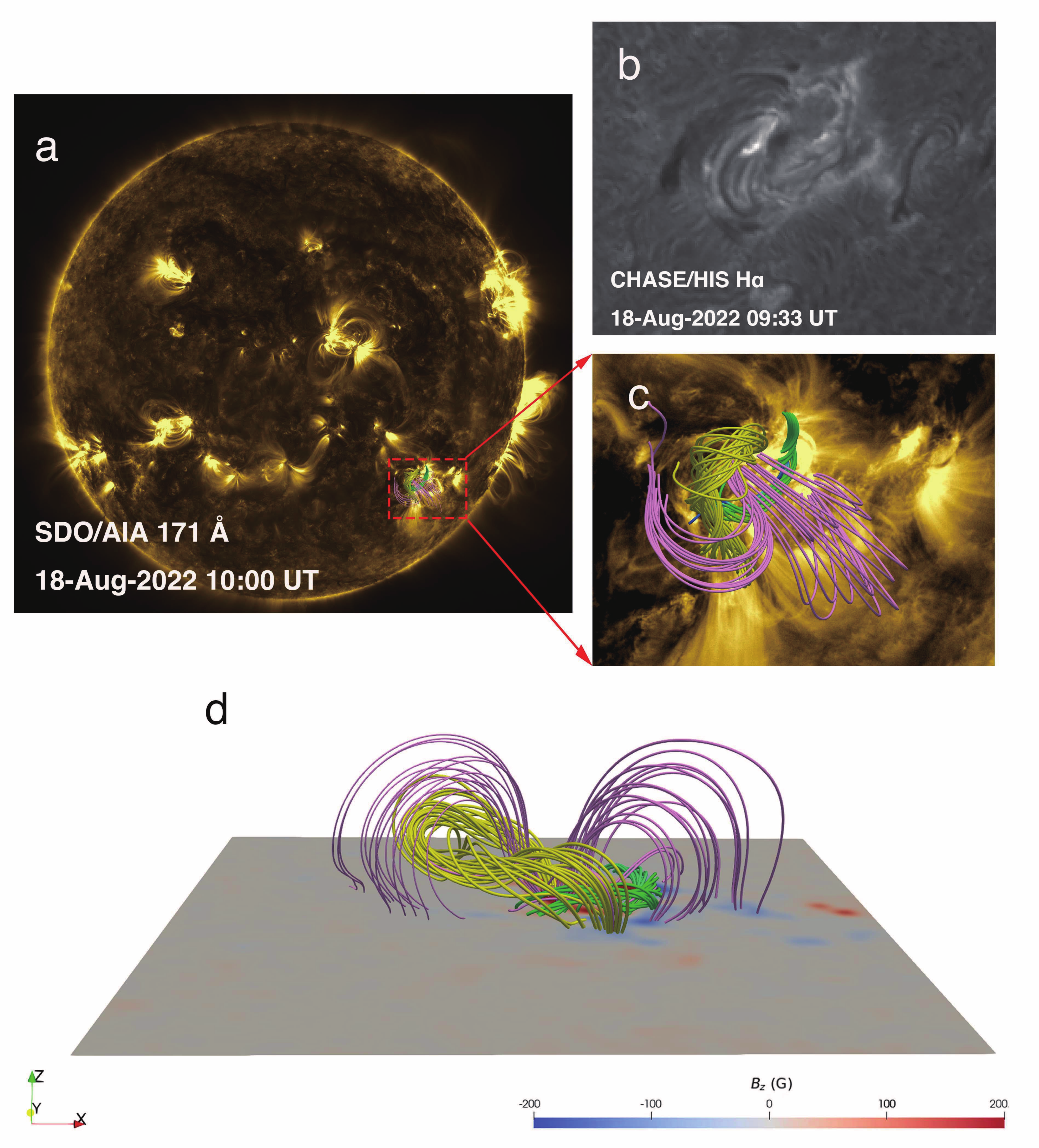}
    \centering
	\caption{Typical magnetic field lines constructed from the NLFFF model at 10:00~UT, which are compared with (a, c) SDO/AIA 171~\AA\ (b) CHASE/HIS H$\alpha$ line-center images. The yellow tubes represent the field lines of the erupting filament, green tubes represent those of the chromosphere fibrils, and pink tubes represent those of the overlying field. (a) Full-disk image of AIA 171~\AA\ at 10:00~UT. Some selected field lines are overlaid in the simulated active region. (b) CHASE/HIS H$\alpha$ line-center images at 09:33~UT. (c) Zoomed-in result of the red dotted box in panel~(a). (d) The side view of the reconstructed magnetic structures. \label{fig6}}
\end{figure*}

The initial magnetic field is provided by the nonlinear force-free field (NLFFF) model, which is constructed with several steps. First, we preprocess the vector magnetogram (hmi.B\_720s) observed by the SDO/HMI to ensure that the photospheric magnetic field satisfies the assumptions of the NLFFF model in the local Cartesian coordinate system. Such a preprocessing involves correcting the projection effects \citep{Guo2017} and removing the Lorentz force and torque \citep{Wiegelmann2004}. Next, we extrapolate the potential field using Green's function \citep{chiu1977} with the $B_{z}$ component and then superpose a flux rope constructed with the Regularized Biot-Savart laws \citep[RBSLs;][]{Titov2018} onto it, where the flux rope path is determined by the filament observed by CHASE/HIS H$\alpha$ line-center image, as shown in Figure~\ref{fig1}c. To maintain the photospheric magnetic field unchanged after inserting the flux rope, as done in our previous works \citep{Guo2019, Guojh2021a}, we subtract the photospheric magnetic fields produced by the RBSL flux rope from the observed $B_z$ before extrapolating the potential fields. As a result, the superposed magnetic fields on the photosphere, which combines the potential field and the flux-rope magnetic field, are consistent with the observational magnetic fields at the solar surface. The RBSL flux rope is mainly controlled by four parameters: the flux rope path ($C$), minor radius ($a$), toroidal flux ($F$), and electric current ($I$). Among them, the first two parameters are approximated as the filament path and width, respectively, while the toroidal flux is estimated as the average of the unsigned flux around the two footpoints of the filament ($F_{0}$), and the electric current is calculated by the equilibrium condition \citep{Titov2018}. After some numerical tests, we select $F=1.5F_{0}=2.5 \times 10^{21}\;$Mx and $a=12.5\;$Mm. The helicity sign is determined to be positive based on the following facts: 1)~the axial magnetic field along the filament spine points to the left when observed from the positive polarity side, corresponding to the sinistral chirality and positive helicity \citep{Martin1998}; 2)~the conjugate drainage sites are right-skewed with respect to the polarity inversion line (PIL), suggesting that the filament is sinistral in chirality \citep{Chen2014}; 3)~this filament is in the southern hemisphere, favoring the sinistral chirality and positive helicity according to the hemispheric preference \citep{Ouyang2017}. Finally, we relax the above superposed magnetic field to a force-free state with the magnetofrictional method \citep{Guo2016a, Guo2016b}. After relaxation, the force-free metric is $\sigma_{J}=0.275$ \citep[see][for details of this metric]{Wheatland2000}. Figure~\ref{fig6} shows some typical field lines of the final NLFFF model, where the CHASE/HIS H$\alpha$ central line and SDO/AIA 171~\AA\ observations are also displayed for comparison. The 3D vector magnetic fields in the simulation are back-projected to the viewing angle of SDO/AIA, which is achieved by employing a matrix composed of three elementary rotations \citep{Guo2017}. It is seen that the twisted flux rope and sheared field lines resemble the observed filament and chromosphere fibrils fairly well, indicating that the NLFFF model is reasonable to serve as the initial condition for the MHD simulation. Regarding the initial density and pressure, we adopt a hydrostatic atmosphere from the chromosphere to the corona, which is described as follows:
\begin{eqnarray}
&&T(z)= \\
&& \begin{cases} T_{_{\rm ch}}+\frac{1}{2}(T_{_{\rm co}}-T_{_{\rm ch}})({\rm tanh}(\frac{z-h_{_{\rm tr}}- 0.27}{w_{\rm tr}})+1)\qquad& z \leq h_{_{tr}},\nonumber \\ (\frac{7}{2}\frac{F_{\rm c}}{\kappa}(z-h_{\rm tr})+T_{_{\rm tr}}^{7/2})^{2/7} & z > h_{_{tr}}, \end{cases}
 \end{eqnarray} \label{eq6}
where $T_{_{\rm ch}}=8000\;$K represents the chromospheric temperature, $T_{_{\rm co}}=1.5\;$MK represents the coronal temperature, $h_{_{\rm tr}}=2\ $Mm and  $w_{_{\rm tr}}=0.2\;$Mm control the height and thickness of initial transition region, and $F_{\rm c}=2 \ \times 10^{5}\ \rm erg \ cm^{-2}\ s^{-1}$ is the constant thermal conduction flux. Hereafter, the density distribution can be calculated from the bottom where the number density is $1.15 \times 10^{15}\rm \ cm^{-3}$.

The data-constrained boundary is assigned to control the photospheric evolution, following the method described by \citet{Guoy2021}. Concretely, we set the velocity in the two bottom ghost layers to zero, and the pressure/density to their initial values, respectively. The vector magnetic fields at the inner ghost layer are fixed to the vector magnetogram observed at 10:00~UT on 2022 August 18, and those at the outer ghost layers are provided by zero-gradient extrapolation. For the remaining five boundaries, the magnetic fields are provided by zero-gradient extrapolation. The density, pressure, and velocities on the side boundary are provided by constant extrapolation, while those on the top boundary are flexible according to the gravitational stratification. Due to the limitation of the Courant-Friedrichs-Lewy (CFL) number, the time step decreases as the $\rm Alfv\acute{e}n$ speed increases. Consequently, simulations involving strong magnetic fields require a longer computation time compared to those with weaker magnetic fields. Additionally, the strong magnetic field in simulations is prone to lead to issues such as magnetic-field divergence and the occurrence of the ``negative pressure'' problem. Therefore, to speed up the computation and reduce numerical dissipation, the magnetic field is modified to one-tenth of the original observed data, see also \citep{Jiang2016, Kaneko2021, Guo2023}. Even so, the evolution of key magnetic structures during the eruption can be captured.

The governing equations (\ref{eq1})--(\ref{eq4}) are numerically solved using the Message Passing Interface Adaptive Mesh Refinement Versatile Advection Code \citep[MPI-AMRVAC\footnote{http://amrvac.org},][]{Xia2018,  Keppens2023}, with a three-step Runge-Kutta time discretization, HLL Riemann solver, and Cada3 limiter. The computational domain is $[x_{min},x_{max}] \times [y_{min},y_{max}] \times [z_{min},z_{max}] = [-157.2, 157.2] \times [-143.0,143.0] \times[1,286.9]\;$Mm, with a uniform grid of $220 \times 200 \times 200$ cells. Our simulation almost covers the entire eruption process from 10:00 to 11:00~UT in the observation.

\section{Results}\label{sec:res}

\begin{figure*}
	\includegraphics[width=13cm,clip]{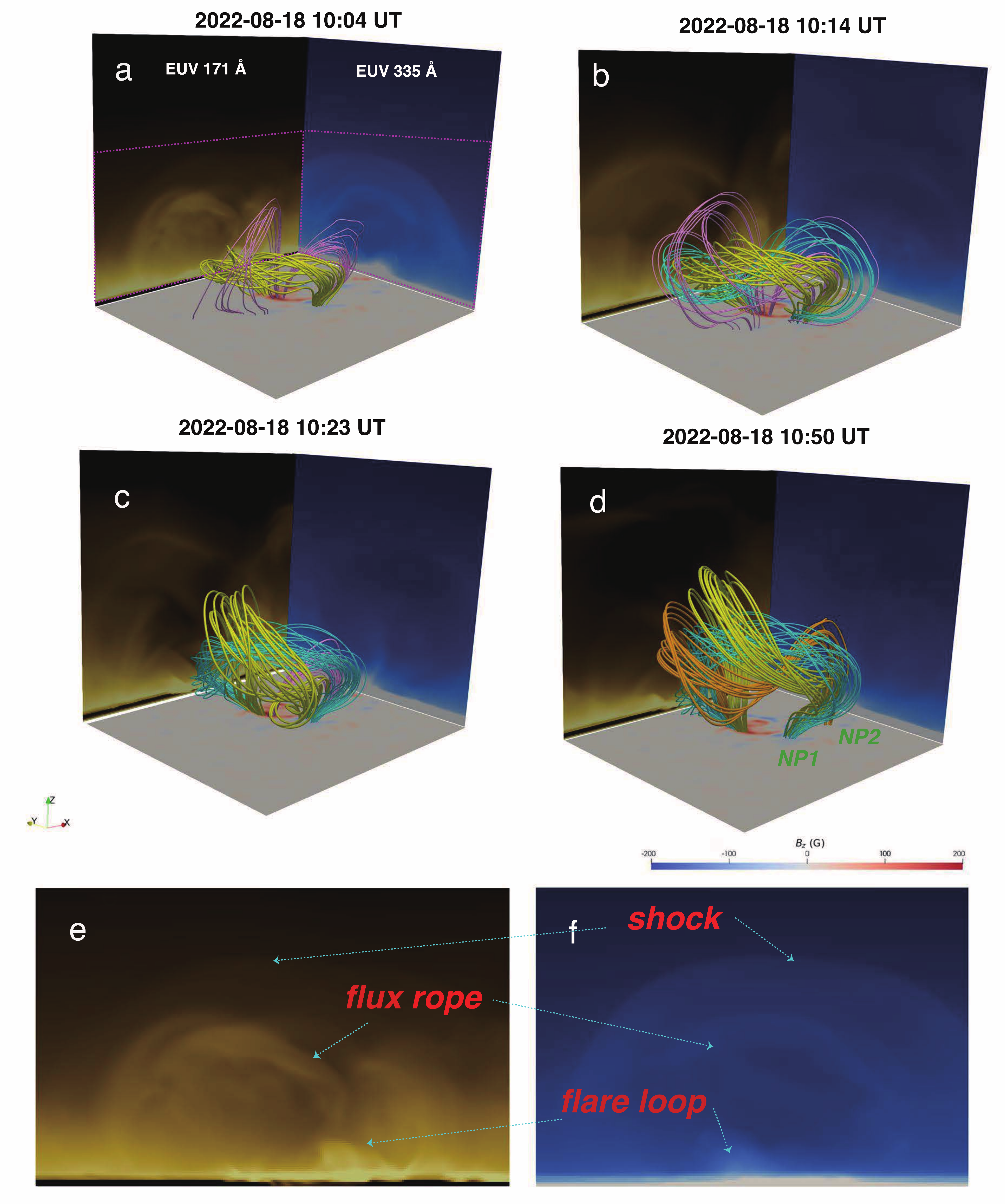}
	\centering
	\caption{Snapshots of the erupting flux rope viewed from the side at (a) 10:04~UT, (b) 10:14~UT, (c) 10:23~UT, and (d) 10:50~UT. Field lines with different topology connectivity are distinguished in color, where the yellow shows the initially inserted flux rope, the cyan and orange show the newly-formed flux ropes with magnetic reconnection, and the pink indicates the background field. Side images in $x$-$z$ and $y$-$z$ planes represent the synthesized EUV images in 171~\AA\ and 335~\AA\ integrated along the $y$- and $x$-axes, respectively. panels~(e) and (f) display the synthesized EUV 171 and 335~\AA\ images with the zoomed-in views of the pink rectangles in panel~(a). \label{fig7}}
\end{figure*}

Figure~\ref{fig7} shows the 3D dynamic evolution of the eruption process viewed from the side. As the flux rope (yellow lines) rises, several overlying and ambient sheared field lines are stretched out, resulting in the formation of a typical hyperbolic flux tube \citep[HFT;][]{Titov2002} structure underneath the flux rope, as depicted in Figure~\ref{fig7}b. HFT is a magnetic topological structure characterized by the intersection of two magnetic separatrix surfaces, providing a favorable setting for the formation of a strong current sheet. Such a topology induces magnetic reconnection between pairs of arcades (pink lines), evolving arcades into twisted field lines (cyan lines) that encircle the original flux rope above the reconnection site and forming flare loops below the reconnection site. With the continuous rising and growth of the erupting flux rope, it encounters adjacent anti-parallel sheared arcades, leading to reconnection, which erodes the original flux-rope leg and causes the westward drifting of the southern footpoint from NP1 to NP2 (Figure~\ref{fig7}d). In this process, some sheared arcades are converted into a part of the flux rope (orange lines). To see the radiative features of this eruption process, we also synthesize the optically thin radiation images at 171 and 335~\AA\ wavebands, as done in \citet{Guo2023}. First, we calculate the emission in each cell with the following formula: $I_{\lambda}(x,y,z)=G_{\lambda}(T)n_{e}^{2}(x, y, z)$, where $G_{\lambda}(T)$ represents the response function for different EUV wavebands. Then we integrate the emissions along the $x$- and $y$-axes to obtain the synthesized images, as illustrated in the two slices in Figure~\ref{fig7}. The synthesized images display several typical eruptive phenomena in observations, including a helical flux rope, a forward shock and an underlying bright flare ribbons, as shown in Figures~\ref{fig7}e and \ref{fig7}f. It should be noted that, the forward shock identified in the simulation is not prominently discernible in observations. This could be attributed to the fact that the initial magnetic fields derived from the NLFFF extrapolation do not precisely adhere to the force-free condition in practice ($\sigma_{J}=0.275$). As a consequence, the residual Lorentz force could expedite the initial ascent of the flux rope compared to observations, thereby favoring the generation of a strong forward shock.

\begin{figure*}
	\includegraphics[width=13cm,clip]{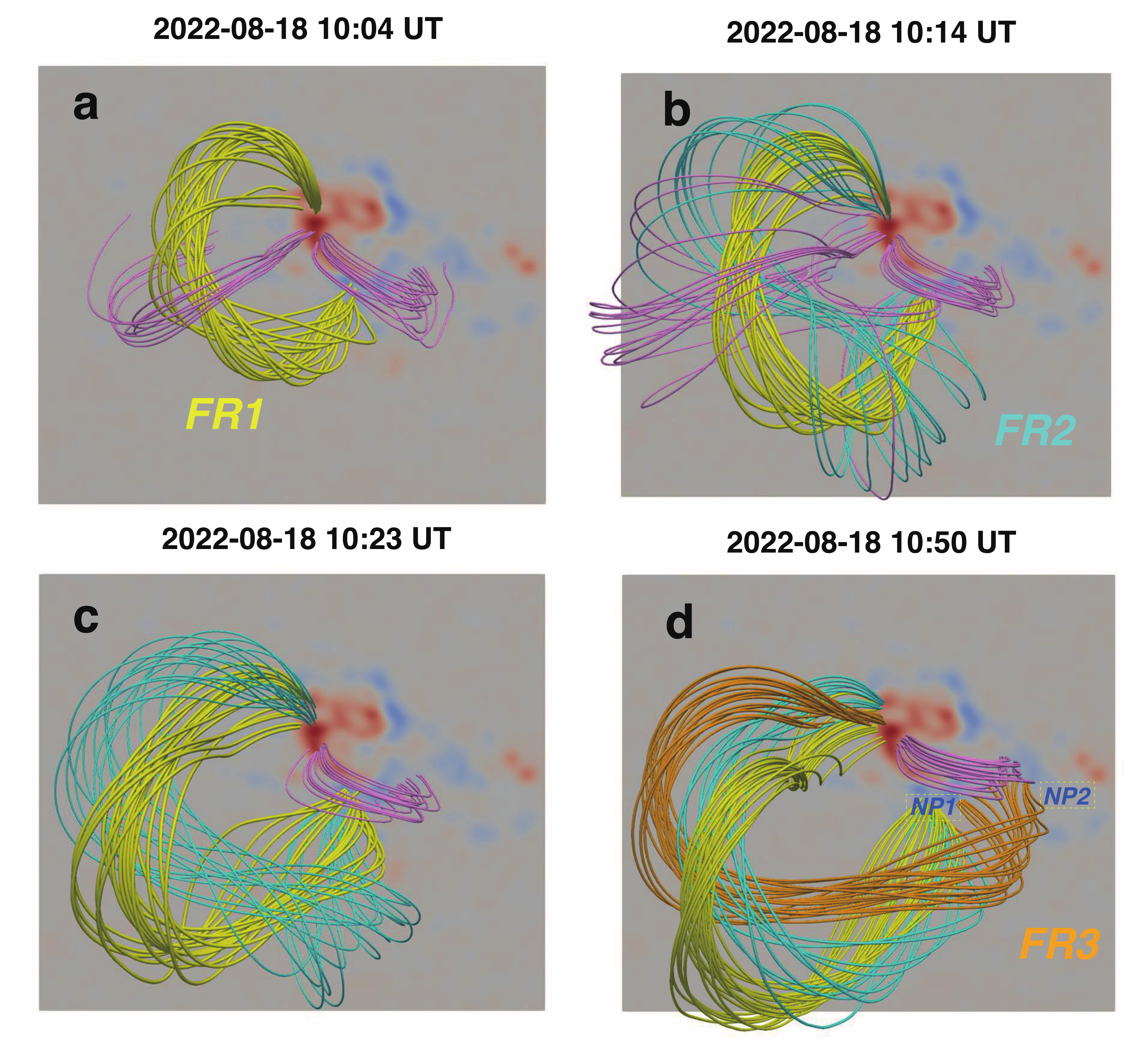}
	\centering
	\caption{Same as Figure~\ref{fig7} but for the top view.  \label{fig8}}
\end{figure*}

Figure~\ref{fig8} displays the evolution of the flux rope viewed from the top. At 10:04~UT, twisted flux tubes, as indicated by the yellow lines, are visible only in the initially inserted flux rope, i.e., FR1. As magnetic reconnection happens below the flux rope as described in the standard CME/flare model, more and more twisted field lines are produced to encircle FR1, forming the envelope of a thicker flux rope, which is labeled as FR2 (cyan lines) in panel~(b). In principle, FR1 and FR2 can be considered as different layers of a single flux rope. As the bulging flux rope erupts, some field lines in the FR1 and FR2 branches may reconnect with the sheared field lines color coded in pink at their southwestern legs. Such interchange reconnection reroutes the southern leg of the former flux rope to further west, forming a separate flux rope which is labeled FR3, as indicated by the golden field lines in panel~(d). Therefore, the eruption of an initial simple flux rope leads to a complex flux rope with three branches at 10:50~UT due to magnetic reconnection in the active region. Whereas FR1 and FR2, albeit their different twists, can be considered as one coherent entity of flux rope as they share common magnetic connectivity, such as the footpoints and the axial directions of the flux ropes, FR3 is apparently distinct from the other two. 

\begin{figure*}
	\includegraphics[width=18cm,clip]{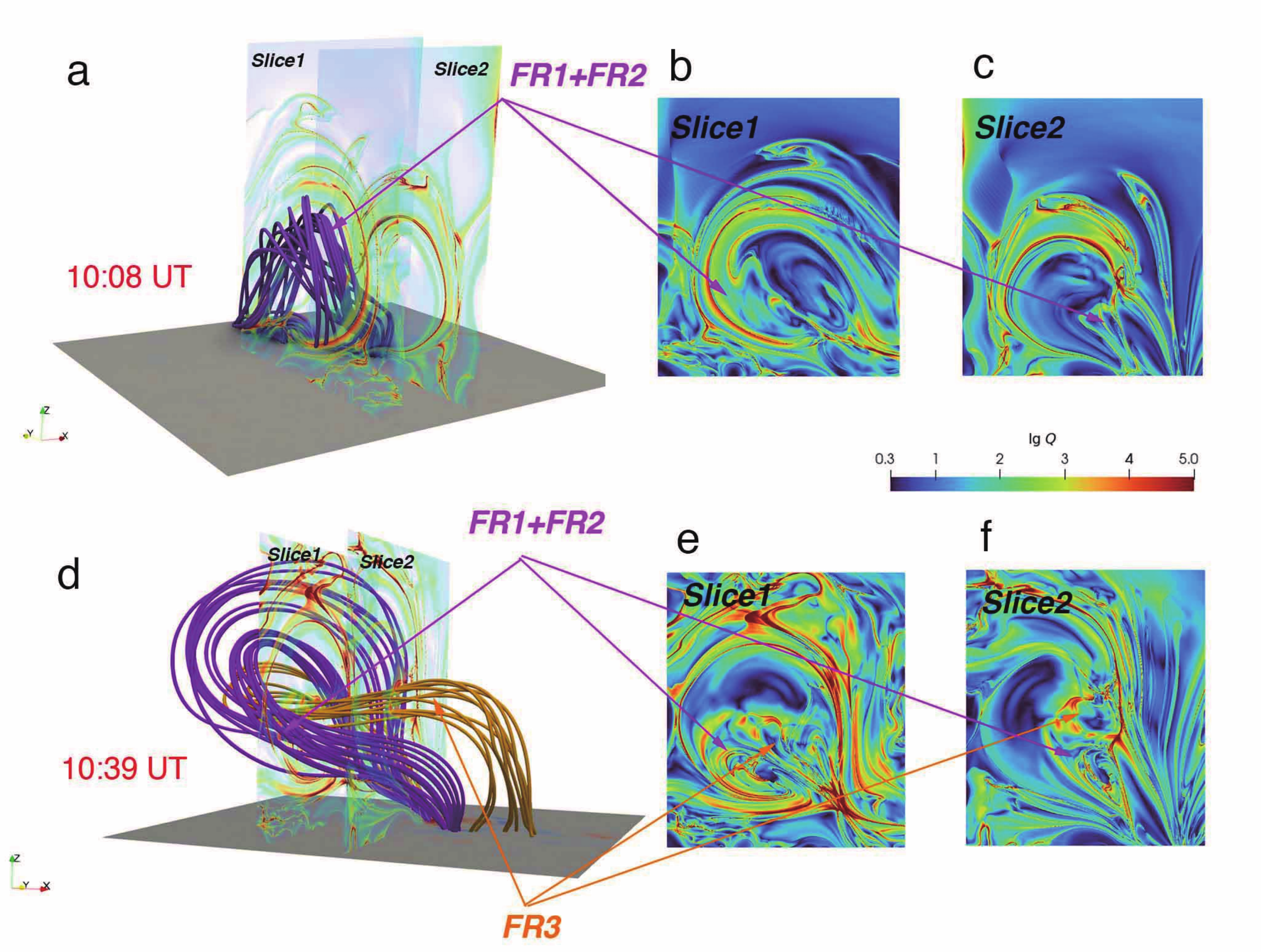}
	\centering
	\caption{Side views of the typical flux-rope filed lines, and the distributions of the squashing factor ($Q$) on the planes of $x=-44\;$Mm and $x=15\;$Mm at 10:08~UT (panels a--c) and 10:39~UT (panels d--f).  \label{fig9}}
\end{figure*}

Quasi-separatrix layers (QSLs) are generally defined as regions where the squashing factor $Q\gg 2$ \citep{Priest1995, demo96, Titov2002}. They are useful in characterizing flux rope boundaries and identifying areas that favor magnetic reconnection \citep{Guo2017, Guojh2021a, Guo2023}. Figure~\ref{fig9} exhibits the typical flux-rope field lines and the distributions of the QSLs, which is computed using an open-source code \citep{Liu2016, Zhang2022}. At 10:08~UT, it is seen that the flux rope consists mainly of FR1 and FR2, which is enclosed by one nearly closed QSL structure. However, at 10:39~UT, we can distinguish two separate branches in the QSL distributions that envelop the FR1/FR2 and FR3 branches, respectively. This indicates that a series of magnetic reconnection in this eruption tends to form an incoherent flux-rope system, transferring it from a coaxial topology to a structure branching into different magnetic domains.

\begin{figure*}
	\includegraphics[width=18cm,clip]{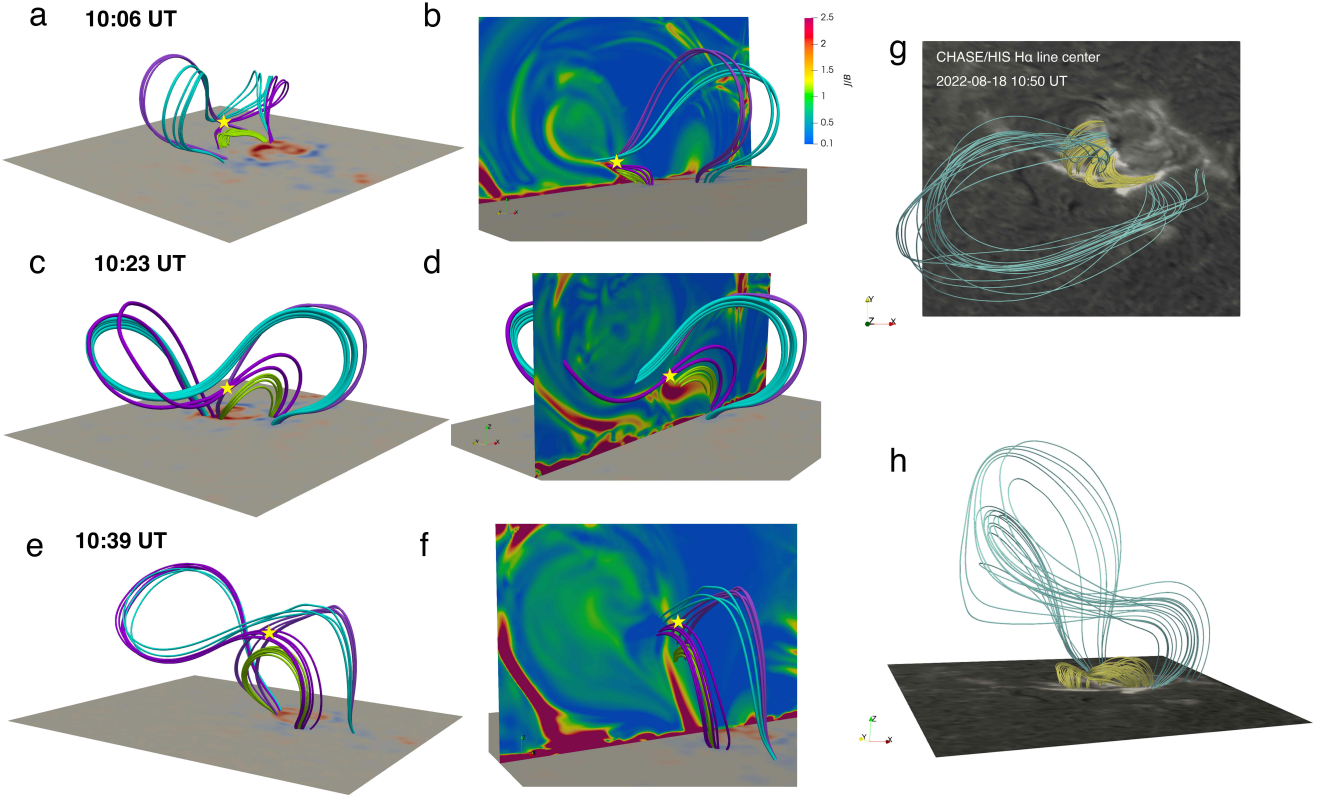}
	\centering
	\caption{panels~(a)--(f) show the magnetic reconnection illustrations recognized by the field-line connectivity (panels a, c and e) and $J/B$ distributions (panels b, d and f) in the $x$-$z$ plane. The purple, cyan, and green lines represent the pre-reconnection field lines, newly-formed flux rope, and flare loops, respectively. The asterisk signs indicate the occurrence sites of magnetic reconnection. Panels~(g) and (h) show the top and side views of the flare loops (yellow tubes) and the simulated flux rope (cyan tubes) at 10:50~UT, respectively. These field lines are overlaid on the CHASE/HIS H$\alpha$ line-center images. \label{fig10}}
\end{figure*}

To further illustrate various reconnection geometries in this event, we present the $J/B$ distributions and the connectivity of some typical field lines in Figure~\ref{fig10}. The scalar metric $J/B$ reflects drastic changes in field-line connectivity, which is often employed to locate the current sheet that favors the occurrence of magnetic reconnection \citep{Gibson2006, Fan2007, Jiang2016, Guo2023}. At 10:06~UT in panels~(a--b), we find two groups of sheared arcades nearly crossing each other (purple lines) around the high $J/B$ region as indicated by the yellow asterisk sign, forming a typical HFT structure. Post-reconnection structures, including a twisted flux rope (cyan lines) and underlying flaring loops (green lines), are also detected. Subsequently, an analogous reconnection geometry is recognized at 10:23~UT but for the different locations of post-reconnection loops, shown in panels~(c--d). Finally, we recognize another reconnection geometry at 10:39~UT in panels~(e--f), namely, the reconnection between the rising flux rope and ambient arcades at the location, which causes the drastic displacement of the flux rope footpoint. As a result, a separate flux rope tied to a further magnetic element with the same polarity is formed. Figures~\ref{fig10}g and \ref{fig10}h present a comparison between the flare ribbons observed by the CHASE/HIS telescope and post-flare loops that are formed as a result of the magnetic reconnection. It is seen that the post-flare loops in the simulation are roughly cospatial with the H$\alpha$ ribbons. This strong spatial correlation suggests that these reconnection geometries may be mainly responsible for the formation of multiple ribbons in the observations.

\begin{figure}
	\includegraphics[width=9cm,clip]{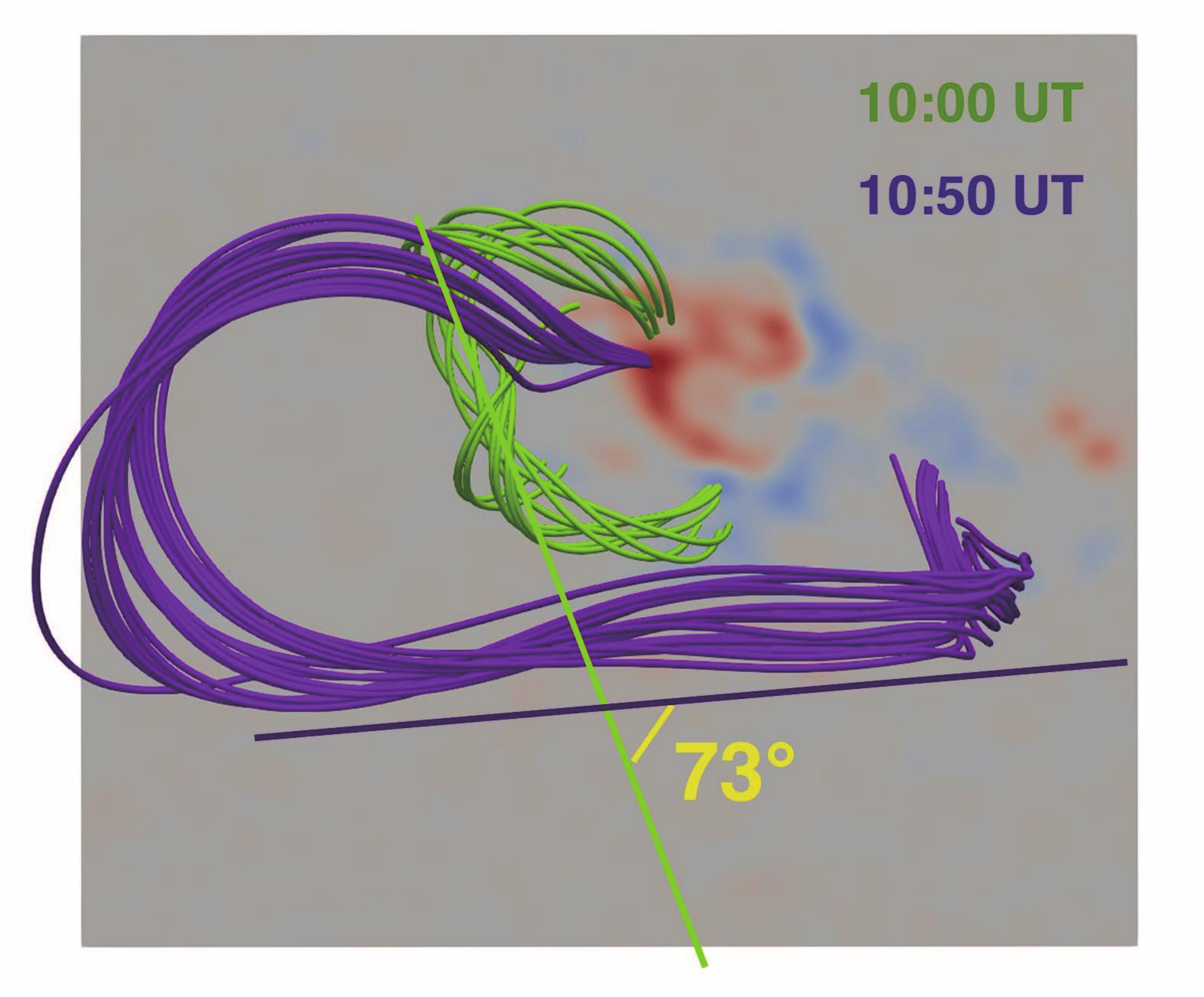}
	\caption{Comparison between the initial flux rope (green) at 10:00~UT and that after reconnection (purple) at 10:50~UT. \label{fig11}}
\end{figure}

Figure~\ref{fig11} depicts the change of the orientation of the flux rope axis resulting from the magnetic reconnection between the erupting flux and the surrounding arcades. Prior to the eruption at 10:00~UT, the field lines of the flux rope are almost aligned from north to south as indicated by the green lines. However, after reconnection at 10:50~UT, they change to an east-west direction. It implies that this reconnection geometry induces a significant change in the axis direction of the flux rope, approximately by $73^{\circ}$. As the reconnection continues, the original flux rope would be eroded continuously, resulting in the ultimate CME flux rope system being predominately composed of the FR3 branch, which leads to the deformation of the CME flux-rope axis. Therefore, we suggest that the magnetic reconnection between the flux-rope leg and ambient arcades is the primary cause for the lateral drifting of the filament material and the later significant deviation of the CME flux rope axis orientation from the pre-eruptive filament in the source region.

\begin{figure*}
	\includegraphics[width=18cm,clip]{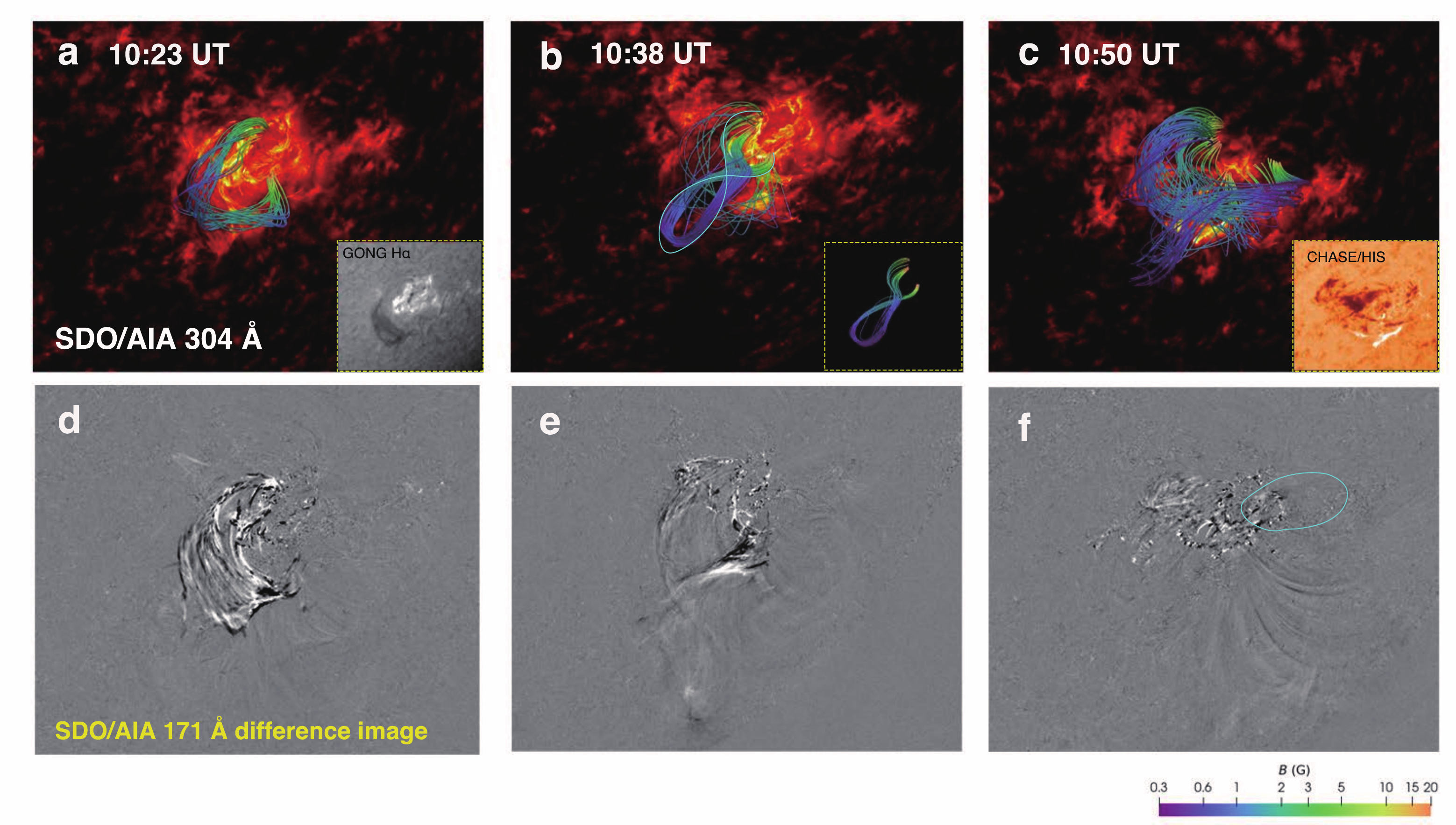}
	\centering
	\caption{Comparison between the simulated flux ropes and observed filaments at (a, d) 10:23~UT, (b, e) 10:38~UT and (c, f) 10:50. panels~(a)--(c) depict the magnetic structures that are back-projected onto the SDO/AIA 304~\AA\ images. The insert figures in panel~(a), (b) and (c) display the GONG/H$\alpha$ line-center image, field lines in a $\gamma$-shaped morphology, and CHASE/HIS H$\alpha$ red-wing image, respectively. panels~(d)--(f) show the SDO/AIA 171~\AA\ running-difference images. The cyan oval in panel~(f) represents the filament material that moves westward.
\label{fig12}}
\end{figure*}

\begin{figure*}
	\includegraphics[width=15cm,clip]{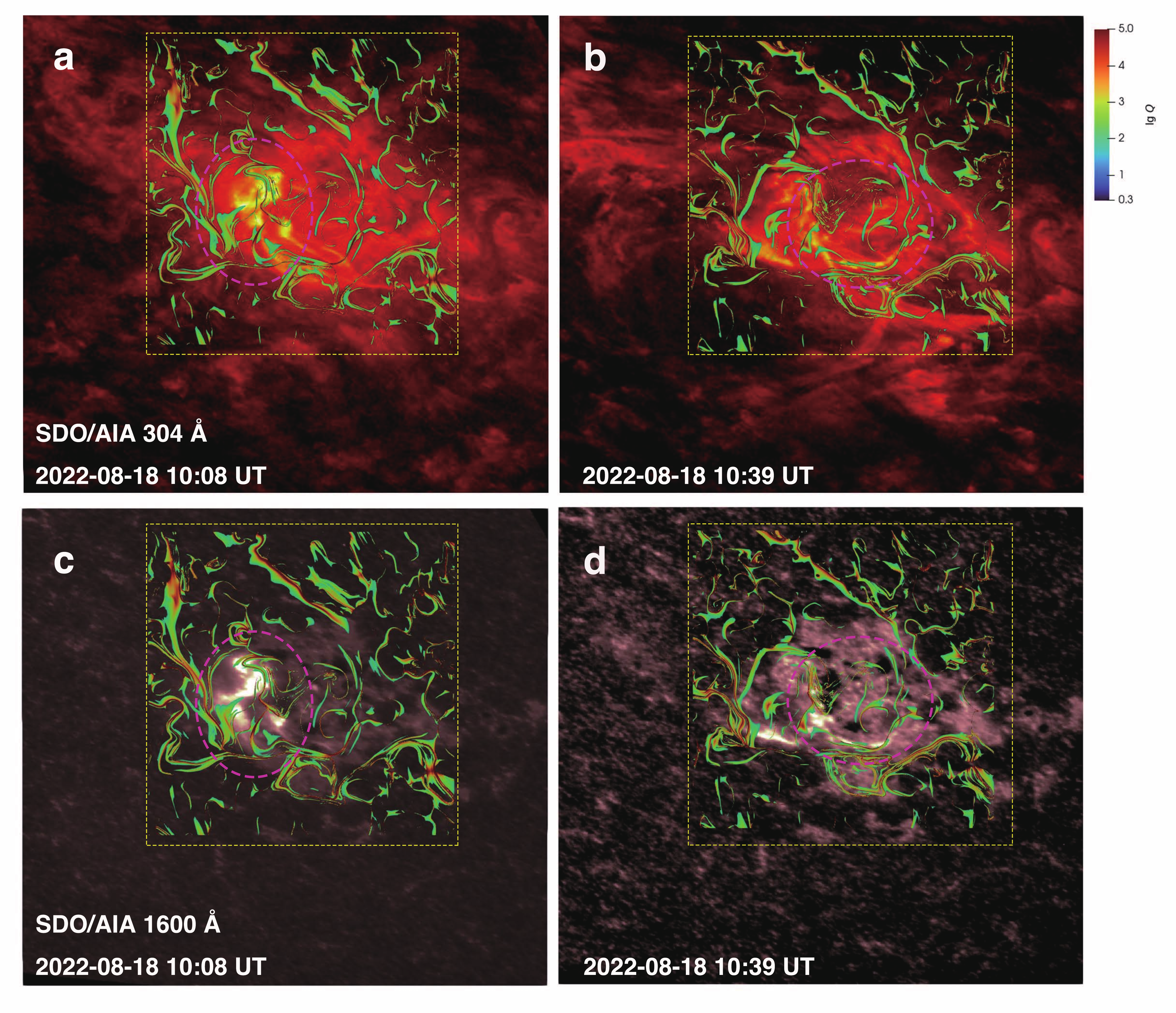}
	\centering
	\caption{Comparison between the simulated QSLs at the solar surface and the EUV images at two moments. The top row is for SDO/AIA 304~\AA\ and bottom row for 1600 \AA. The left column is for 10:08~UT and the right column for 10:39~UT. The pink ovals highlight the QSLs corresponding to the flare ribbons.  \label{fig13}}
\end{figure*}

To validate the reliability and capability of our MHD modeling in capturing the key physical processes in observations, we compare the simulated flux rope with the observed filament, as depicted in Figure~\ref{fig12}. As discussed in \citet{Chen2020}, the temperature of the filament material could undergo an increase owing to the enhanced coronal heating during the eruption, which may lead to changes in visible wavelengths and radiation characteristics (emission or absorption) of the eruptive filament. This phenomenon has been found in both the observations conducted by \citet{Lee2017} and a thermodynamic MHD simulation that incorporates an eruptive filament \citep{Guo2023}. As such, to track the evolution of the filament during the eruption, we utilize the observations from SDO/AIA, CHASE/HIS, and Global Oscillation Network Group (GONG)/H$\alpha$ \citep{Harvey2011}. Specifically, we employ EUV 304~\AA\ and H$\alpha$ line-center wavebands to observe the relatively cold filament material, EUV 171~\AA\ to trace the heated filament material during the eruption, and H$\alpha$ red-wing to reflect the material undergoing falling and lateral drifting motions. It is found that the morphology of the simulated flux rope resembles that of the observed filament quite well. Moreover, the eruption path of the simulated flux rope leans toward the southeast, which is in line with the observed filament. At 10:23~UT, both simulated flux rope and the observed filament exhibit a semi-circular shape, which both evolve into a $\gamma$ shape at 10:38~UT. During this stage, the orientations of the erupting filament and flux rope are roughly along the north-south. However, at 10:50~UT, there is an emergence of some flux-rope field lines exhibiting the east-west orientation. These field lines may serve as pathways along which the filament material moves, leading to the manifestation of lateral drifting motion. To further validate the rationality of the magnetic topology evolution in the simulation, we compute the squashing factor in the photosphere, as shown in Figure~\ref{fig13}. The QSLs almost retrieve the prominent features in observations. One can see that the shape of the brightening regions matches the QSLs very well, which are highlighted by pink ovals.

\section{Discussions and summary} \label{sec:dis}

In this paper, we conduct a data-constrained MHD simulation for a filament eruption that occurs on 2022 August 18. The filament material exhibits evident lateral drifting during its ascent as observed by both SDO/AIA and CHASE/HIS. The comparison between the observed filament direction in the source region and the GCS reconstruction for its ensuing CME suggests that the flux-rope axis deforms during the propagation from the solar surface to the interplanetary space. Our simulation successfully reproduces several features in observations. For example, the simulated flux rope erupts initially towards the southeast with a semi-circle shape, which is in line with the observed erupting filament. Additionally, the simulated flux-rope axis undergoes a significant detour from its initial state, as found in observations. The matching between the simulation and the observations provides an opportunity to decipher the physics behind it.

\subsection{What leads to the filament lateral drifting and CME flux rope departure from the pre-eruptive filament?} \label{sec:le}

It has been revealed that frequently the magnetic-field orientation of ICMEs matches that of the pre-eruptive filaments in the source region \citep{Bothmer1994, Yurchyshyn2001}. Such a relationship allows us to infer the ICME magnetic structure near the Earth from the pre-eruptive filament in the low corona. However, from time to time, the direction of the ICME flux rope axis observed near the Earth was found to deviate from that of the pre-eruptive filament \citep{Wang2006, Liuy2018}. This means that CMEs do not always keep their initial orientation and may experience deformation when traveling from the corona to the interplanetary space. Both the remote sensing and in-situ observations have found that not all halo CMEs originating from the filament eruption near the solar disk center can eventually hit the Earth, and some CMEs that erupt from the solar limb can still reach and influence our planet \citep{Wang2002, Gopalswamy2009, Wang2014, Shen2022}. Moreover, the interactions between CMEs may also lead to its magnetic-field rotation, resulting in the formation of geoeffective CMEs \citep{Maharana2023}. These would confuse our judgment in determining whether the CMEs can reach the Earth or not, and assessing their potential geomagnetic effects. 

Numerous studies have demonstrated that the interactions between CMEs and the solar wind in the interplanetary space can result in the deflection of CME flux ropes. For instance, \citet{Wang2004} discovered that fast ICMEs are blocked by slow solar wind streamers ahead of them, causing them to deflect eastward. On the other hand, slow ICMEs may be propelled by trailing fast solar wind streamers behind them, leading to the deflection in the western direction. This scenario has been supported by numerical simulations performed by \citet{Zhuang2019}, who pointed out that the velocity difference between CMEs and the background solar wind is a critical factor in determining their deflection direction. 

Compared to the physical processes in the interplanetary space, the magnetic field in the solar corona is more intricate and powerful, opening up many more possibilities for inducing the rotation and deflection of a flux rope. As for the rotation, kink instability \citep{Torok2004} and shear-field components \citep{Kliem2012} are the major drivers. Apart from these, interchange reconnection between the closed flux rope and adjacent open fields can also alter the magnetic structure \citep{Lugaz2011, Cohen2011}. In particular, the global MHD simulations conducted by \citet{Shiota2010} demonstrated that the magnetic reconnection between the flux rope and ambient field lines can result in the rotation of the CME flux rope. As for the deflection, \citet{Shen2011} and \citet{Gui2011} pointed out that the gradient of magnetic field can alter the trajectory of a CME from its original path. Such structures that would lead to CME deflections are widespread in the solar corona, including weak magnetic regions like the heliospheric current sheet and pseudo streamers, and strong magnetic regions such as active regions, which can guide the deflection of flux ropes \citep{Cremades2004, Wang2011, Karna2021, Mostl2015}. Besides, \citet{Chen2000} found that the emerging flux on one side the flux rope can trigger the flux rope eruption and cause it to lean toward the emerging flux side.

In this paper, we find the other possibility to deform the flux rope structure, which can be explained on the basis of the 3D flare model proposed by \citet{Aulanier2019}. They summarized three magnetic reconnection geometries in the eruption process: 1)~aa-rf reconnection, which occurs in the overlying arcades to form a flux rope and flare loops simultaneously; 2)~ar-rf reconnection, which takes place between the flux rope and neighboring arcades, changing the connectivity of the flux rope; 3)~rr-rf reconnection, which occurs in flux rope field lines, increasing the twist number of a flux rope. Among them, `a' represents the shear arcade, `r' the flux rope, and `f' the flare loop. This 3D model works well in quantifying the flux evolution and footpoint drifting of flux ropes during eruption \citep{Zeman2019, Dudik2019, Xing2020}. Furthermore, a natural consequence of the footpoint displacement of the flux rope is its axis deformation \citep{Jiang2013}, in particular the direction of the arcades involved in the reconnection are quite different from that of the flux rope, which ultimately causes its orientation to deviate from its pre-reconnection state. 

\begin{figure*}
	\includegraphics[width=17cm,clip]{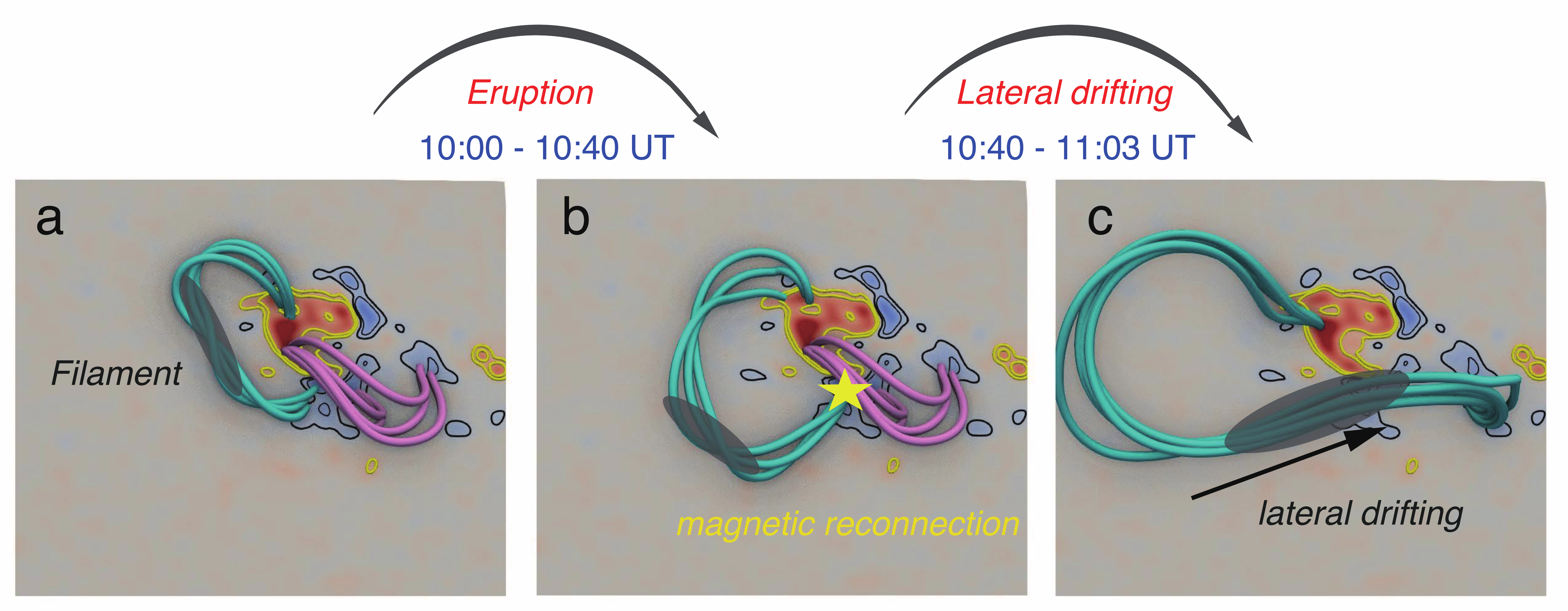}
	\centering
	\caption{Sketch illustrating the eruption and lateral drifting of the filament. The cyan and pink tubes represent the flux rope carrying the filament material and the ambient arcades, respectively. The yellow asterisk locates the regions where interchange magnetic reconnection may occur. The gray translucent contours represent the filament material. \label{fig14}}
\end{figure*}

By analyzing the simulation data, we identify two reconnection geometries, namely, the reconnection in the overlying field lines (aa-rf) , and the reconnection between the flux rope and ambient arcades (ar-rf). The former injects magnetic fluxes and twisted field lines to encircle the original flux rope, whereas the latter significantly causes the displacements of flux-rope footpoints and alter its axis direction. Meanwhile, filament material moves along the continuously reconnected flux rope and naturally exhibits the lateral drifting. A physical scenario explaining the observed phenomena is illustrated with the sketch in Figure~\ref{fig14}. Between 10:00~UT and 10:40~UT, the filament rises and leans toward the southeast, as shown in panels~(a) and (b). Subsequently, interchange reconnection between the original flux-rope field lines that carry the filament material and the ambient arcades occurs. This leads to a change in the axis orientation of the flux rope, causing the filament material to move along the newly formed flux rope and exhibits the lateral drifting motions, as illustrated in panel~(c). Particularly, \citet{Dudik2019} suggested that the shift of the filament leg may be a manifestation of ar-rf reconnection. In this paper, we propose that the lateral drifting of the filament material can also serve as a signature for this 3D reconnection geometry.

Additionally, the reconnection also results in the axis deformation and departure of the CME flux rope from the pre-eruption filament. As depicted in Figure~\ref{fig11}, the post-reconnection field lines deform a lot relative to those prior to eruption. This means that the substantial displacements of the footpoints of the flux rope could completely reconstruct its morphology. Recently, \citet{Gou2023} claimed that the long-distance migration or jump of the footpoint of the flux rope, associating with complete replacement of the flux during the eruption. Based on the observations and MHD simulations presented in this paper, it has been indicated that in addition to the footpoint jump, the lateral drifting of the filament material can also serve as a signature that the flux rope is being reconstructed during the eruption. Taken as a whole, our work suggests that the ar-rf reconnection geometry may play a significant role in redirecting the axis of the flux rope, and the observed lateral drifting of the filament material is likely a manifestations of this process.

\subsection{Filament drainage sites: the hint to reflect flux rope structures in the eruption} \label{sec:le}

Solar filaments are cold and dense plasmas often hosted in magnetic dips, which are commonly modeled as either sheared arcades or twisted flux ropes \citep{Mackay2010, Chen2020}. Filament fine structures and oscillations are well believed to be intimately related to their supporting magnetic structures \citep{Chen2020}. For the former, the length and dynamics of filament threads can indicate the structure of the supporting magnetic dip \citep{Luna2012a, Zhou2014, Guojh2021b, Guo2022}; the filament morphology, including aspect ratio, composition of threads, horn and cavity, can show the twist degree of a flux rope \citep{Guojh2021a, Guo2022}; the filament barbs may indicate the bifurcations of a flux rope due to parasitic polarities in the surrounding environment \citep{Aulanier1998}, and the fibrils in a filament channel can be used to determine the chirality of the filament \citep{Martin1998}. As for the filament oscillation, longitudinal oscillations have been widely applied to derive the curvature radius of local field lines \citep{Luna2012b, Zhang2012, Zhang2013, Zhang2020, Zhou2017, Zhou2018, Ni2022}, and transverse oscillations can be utilized for the estimate of the magnetic field strength around the filament \citep{Shen2014, Zhou2018}. 

Apart from these, there are other observable phenomena in the eruption site that can provide insights into the magnetic structures of erupting filaments. For example, \citet{Chen2014} revealed that some filament materials drain back down to the solar surface during eruption, forming a group of conjugate brightening sites on two sides of the PIL. They noted that the skewness of the drainage sites can indicate the chirality of the filament, i.e., left (right)-skewed drainage sites with respect to the PIL correspond to the dextral (sinistral) filament \citep[see also][]{Wang09}. It is noted that this approach is universal for filaments supported by either sheared arcades or flux ropes, while Martin's rule on filament barbs \citep{Martin1998} is only applicable to the filaments supported by flux ropes, which means that the filament barb chirality is determined by magnetic configuration and helicity simultaneously \citep{Guo2010, Chen2014}. For example, a flux rope with negative helicity would lead to right-bearing barbs, while a sheared arcade with the same helicity leads to left-bearing barbs. Based on this, \citet{Chen2014} proposed a method to determine the magnetic configuration of a filament by observing the skewness of filament barbs or threads and the drainage sites once the filament erupts. Applying this approach to 576 eruptive filaments, \citet{Ouyang2017} found that approximately 89\% of filaments are supported by flux ropes, whereas only 11\% of them are sheared arcades.

The studies mentioned above mainly focused on diagnosing the magnetic structures of filaments before they erupt. However, a crucial question remains: how can we diagnose the magnetic structures of a filament in the eruption? One immediate recipe is to trace the morphology of the erupting filament in real time. However, this may be difficult to achieve in practice due to the enhanced heating, mass drainage and expansion of the flux rope during eruption, which can cause the filament to become more invisible as it rises \citep{Chen2020}. Therefore, real-time tracking could be challenging. As a result, many previous works took a fairly simplistic approach, assuming that the filament spine observed before eruption is the direction of the flux rope axis \citep{Bothmer1994}, and comparing it with the resulting ICME, ignoring the possibility that the filament may undergo deformation during eruption. However, this approach stands a good chance to lead to misjudgments. First, the spatial relationship between the filament and its supporting magnetic flux rope is not straightforward as people thought \citep{Guo2022}. As shown in Figure 6 of \citet{Guo2022}, filament material occupies only a small portion of the flux-rope bulk rather than the entirety of it. Moreover, the ar-rf reconnection geometry can erode the original flux-rope leg and alter it to the side of the arcade. Fortunately, the drainage sites shed some light on the erupting magnetic structures, which can tell us where the flux rope is anchored after the flux rope has been deformed. Therefore, we recommend taking additional factors into consideration when determining the direction of a CME flux rope with its pre-eruptive filament. Specifically, the filament spine is utilized to outline the overall path of the flux rope, and the drainage sites can be used to determine where the flux rope is anchored. By doing so, we can diagnose the magnetic structures of erupting filaments more clearly. Another thing worth noting is that the early evolution of the CME flux rope from the solar surface to 2 solar radii is usually missing in coronagraph observations due to the field-of-view gap, and EUV imaging observations might be able to fill in the gap, for instance, the slow-component EUV wave might be the counterpart of the corresponding CME frontal loop \citep{Chen2016}, which can be used to investigate the CME triggering and acceleration; the drainage sites of erupting a filament reflect the footprints of its supporting flux rope, which can reveal its magnetic configuration and magnetic reconnection processes it possibly experiences.

The CHASE mission provides full-disk H$\alpha$ spectroscopic observations, which have the advantage in deciphering the supporting magnetic structures of filaments, as exhibited in this paper. It is widely accepted that the H$\alpha$ spectral line is optimal for filament observations. With the high-resolution full-disk H$\alpha$ images across the H$\alpha$ profile provided by CHASE/HIS, the fine structures of filaments can be clearly resolved (Figure~\ref{fig1}c), enabling us to diagnose the magnetic fields of filaments, such as the orientation and helicity sign of their supporting flux rope \citep{Chen2020}. In the future, these observational data could also set a stage for examining our previously published results based on numerical simulations. For example, long/short filament threads are prone to be formed in shallow/deep magnetic dips \citep{Zhou2014, Guojh2021b}; the fairly short threads that present decayless oscillations are likely to be hosted in multiple-dipped flux tubes \citep{Zhou2017, Guojh2021b}; short-lived threads forming high-speed flows inside the filament imply that they are likely to be supported by weakly twisted flux tubes, and quasi-stationary threads that generally present the oscillations are more likely to be hosted in highly twisted flux tubes \citep{Guo2022}. In addition, the CHASE spectroscopic observations can also provide full-disk Doppler velocity distribution with the H$\alpha$ spectra. On the one hand, they can help us locate the stealth drainage sites in EUV wavebands, and thus elucidate the erupting magnetic structures. Apart from that, the Doppler velocity fields are crucial in unveiling the nature of some elusive phenomena about filaments, like the vertical threads of some prominences \citep{Schmieder2014} and the solar tornadoes \citep{Schmieder2017, Yang2018, Gunar2023}. Given these findings, we believe that the CHASE observations are valuable in determining CME structures and forecasting their space weather effects.

\acknowledgments
We would like to thank the anonymous referee for his/her fruitful comments which help us to improve the paper. We acknowledge the use of data from CHASE, SDO, STEREO, and SOHO. The CHASE mission is supported by China National Space Administration. This research was supported by National Key R\&D Program of China (2022YFF0503004 and 2020YFC2201200), NSFC (12127901, 11961131002, 11773016 and 11533005). J.H.G. and Y.H.G were supported by the China Scholarship Council under file No.\ 202206190140 and No. 202206010018, respectively. S.P.\ acknowledges support from the projects C14/19/089  (C1 project Internal Funds KU Leuven), G.0B58.23N  (FWO-Vlaanderen), SIDC Data Exploitation (ESA Prodex-12), and Belspo project B2/191/P1/SWiM. The numerical calculations in this paper were performed in the cluster system of the High Performance Computing Center (HPCC) of Nanjing University.

\bibliography{ms}{}
\bibliographystyle{aasjournal}

\end{CJK*}
\end{document}